# A Prototype of LaBr3:Ce in situ Gamma-Ray Spectrometer for Marine Environmental Monitoring

**Ming Zeng[1]**
*Key Laboratory of Particle & Radiation Imaging (Tsinghua University), Ministry of Education*
*Department of Engineering Physics, Tsinghua University, Beijing, 100084, China*
*E-mail:* `zengming@tsinghua.edu.cn`

**Zhi Zeng, Jirong Cang, Xingyu Pan, Tao Xue, Hao Ma, Hongchang Yi, Jianping Cheng**
*Key Laboratory of Particle & Radiation Imaging (Tsinghua University), Ministry of Education*
*Department of Engineering Physics, Tsinghua University, Beijing, 100084, China*
*E-mail:* `zengzhi@tsinghua.edu.cn`

A prototype of LaBr3:Ce in situ gamma-ray spectrometer for marine environmental monitoring is developed and applied for in situ measurement. A 3-inch LaBr3:Ce scintillator is used in the detector, and a digital pulse process electronics is chosen as the pulse height analyzer. For this prototype, the energy response of the spectrometer is linear and the energy resolution of 662keV is 2.6% (much better than NaI). With the measurement of the prototype in a water tank filled with 137Cs, the detect efficiency for 137Cs is (0.288±0.01)cps/(Bq/L), which is close to the result of Monte Carlo simulation, 0.283cps/(Bq/L). With this measurement, the MDAC for 137Cs in one hour has been calculated to 0.78Bq/L, better than that of NaI(Tl) in-situ gamma spectrometer, which is ~1.0Bq/L.

---

[1]   Ming Zeng

## 1. Introduction

NaI(Tl) detector is the main choice in marine environmental monitoring because of its high detection efficiency、stable performance、low price. However the energy resolution of NaI(Tl) detector is not very high, which limits its performance. LaBr$_3$:Ce crystal has a series of advantages including high scintillation efficiency、high energy resolution, which makes LaBr$_3$:Ce a potential good choice despite the self-activity in LaBr$_3$:Ce crystal. The main radionuclides in LaBr3 crystal are $^{138}$La and progenies of $^{227}$Ac. $^{138}$La is a radioisotope of lanthanum and $^{227}$Ac exists as a radioactive contamination in LaBr$_3$ crystal, which is a radionuclide in the natural $^{235}$U decay chain. The preliminary test result of this prototype is introduced in details.

## 2. Framework

This prototype mainly consists of underwater and aquatic two parts (See Fig. 1). Underwater system, which is responsible for the collecting of spectra, is made up of a Saint Gobain 3''×3'' BrilLance-380 Detector, R10233-100 photomultiplier tube (PMT), Hamamatsu CC228 high voltage supplier，Amptek DP5 multichannel analyzer (MCA) etc. Aquatic system is responsible for the power management, using the solar power, and data communication.

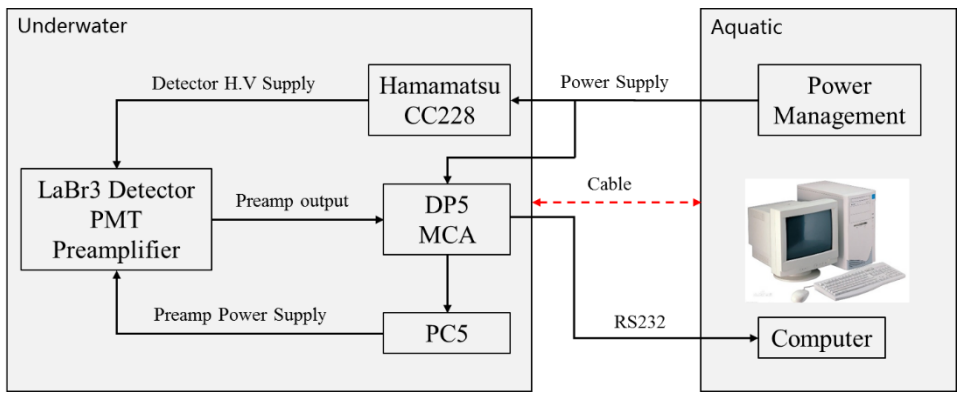

Fig. 1 System Framework

Hardware system and the prototype is shown as Fig.2.

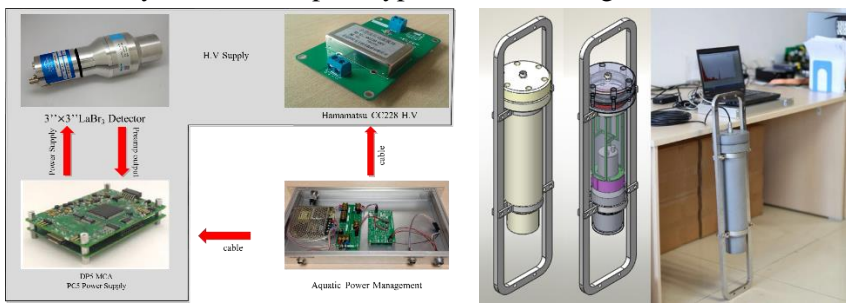

Fig. 2 Hardware Framework and Prototype

## 3. Energy resolution

A set of calibrated radioactive sources were used to characterize the response of the detector at various gamma-ray energies. The complete list of sources used together with their energies are reported in Table 1.

Table 1. Nuclide list used in the calibration measurement

| Nuclide | Energy/keV | Nuclide | Energy/keV |
|---|---|---|---|
| Cs-137 | 661.66 |  | 867.38 |
| Co-60 | 1173.23 | Eu-152 | 1528.1 |
|  | 1332.49 |  | 276.40 |
| Eu-152 | 344.28 | 133Ba | 302.85 |
|  | 411.12 |  | 356.01 |
|  | 443.97 |  | 383.85 |
|  | 778.90 |  |  |

Single-energy photo peaks, without lying the Compton edgy, are chosen. According to the same analysis, we have evaluated the FWHM versus energy.

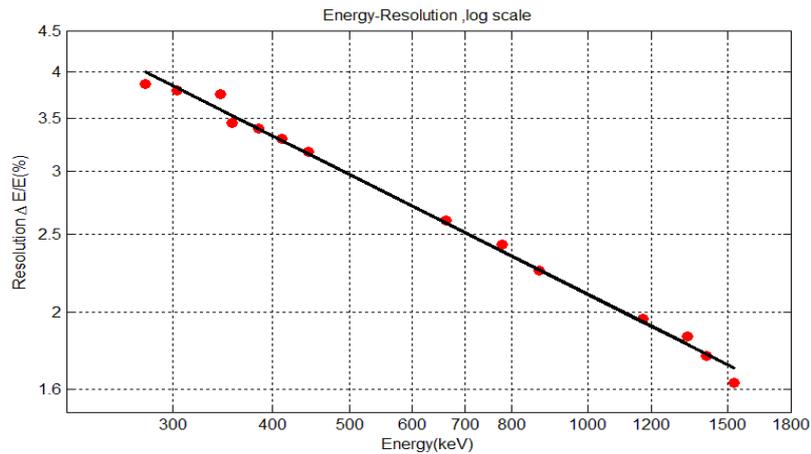

Fig. 3 Energy resolution versus energy

The data are best-fitted by power laws as:

$$\frac{\Delta E}{E}(\%) = 65.46 E^{-0.4975}$$

The exponent of ~0.5 indicates that the noise contributed by electronics is small. According to last equation, the energy resolution is 2.6% @662keV.

## 4.Intrinsic background

One main reason that LaBr3:Ce detector isn't used in low background measurement widely is the intrinsic background of LaBr3. The main radionuclides in LaBr3 crystal are $^{138}$La and progenies of $^{227}$Ac. $^{138}$La is a radioisotope of lanthanum and $^{227}$Ac exists as a radioactive contamination in LaBr3 crystal, which is a radionuclide in the natural $^{235}$U decay chain.

To investigate the intrinsic activity of LaBr3 detector, the background spectrum (shown in Fig.4) of the LaBr3:Ce in-situ gamma-ray spectrometer was measured in lead shield. From the spectrum, we can distinguish several energy peaks including the 35.5keV X ray、255 keV β continuous spectrum、the combination of 789 keV γ ray and 255 keV β continuous spectrum、the 1436 keV γ ray、the combination of 1436 keV γ ray and 35.5 keV X ray, which come from the radioactive decay of $^{138}$La in LaBr3 crystal. Besides, continuous spectrum from 1600 keV to 3000 keV result from the α decay of progenies from $^{227}$Ac. The total counting rate of the background is about 1.08cps/cm$^3$ for the 3''×3'' LaBr3 detector. And, the counting rate of the α contamination occurring in the energy region from 1600keV to 3000keV is about 0.09cps/cm$^3$.

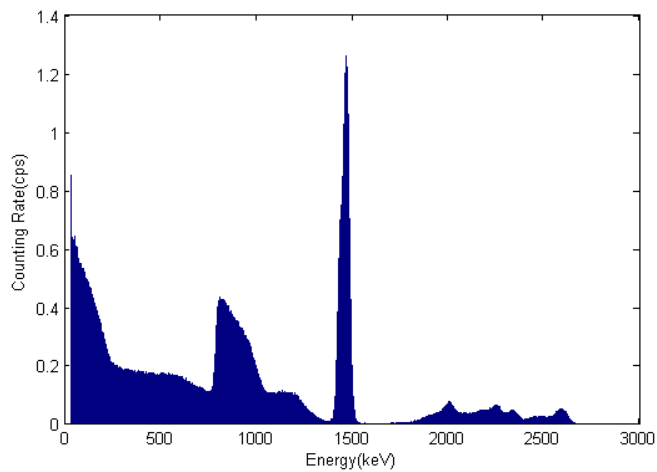

Fig. 4 Intrinsic background for 3''  3'' LaBr3

## 5.Detection efficiency

We define detection efficiency $I_\gamma \, \varepsilon_{sp} \cdot V$, which is a sufficient parameter for the prototype, as the photo peak count rate when the activity concentration of $^{137}$Cs is 1Bq/L, where $I_\gamma$ is the absolute intensity of photons, $\varepsilon_{sp}$ is ratio between photons number and counts recorded in the full energy peak, V is the volume of the area containing particle source.

We both simulated and measured the detection efficiency of the prototype in a seawater tank. The seawater tank is $\Phi$2m×2.3m containing 0.48Bq/L $^{137}$Cs radioactive source.

The simulation was done with Geant4. And the geometrical model of seawater tank and the detector is shown in Fig.5. Considering the influence of distance between particle source and LaBr3 crystal, we simulated the detection efficiency of the prototype to $^{137}$Cs in seawater sphere with different radiuses. The relationship between radius and detection efficiency indicates that the detection efficiency of this prototype is about 0.283cps/(Bq/L) when the area of seawater is large enough, which means the sphere is about 80cm.

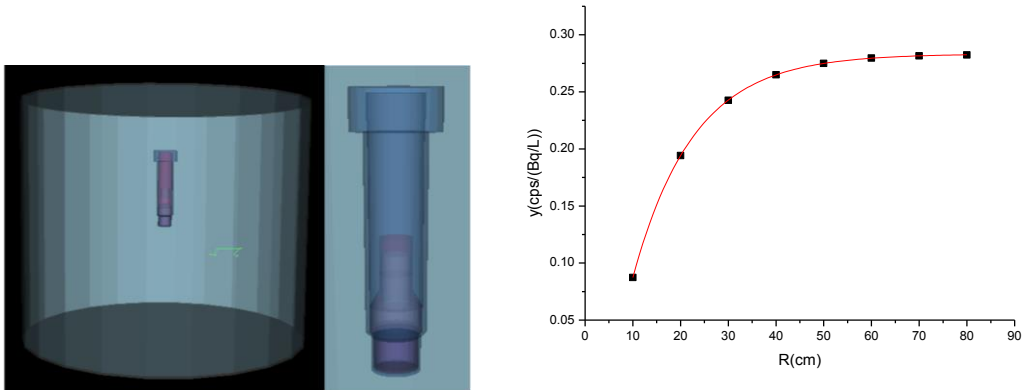

Fig. 5 Model and result of the detection efficiency simulation

A spectrum was measureed actually in the seawater tank, the photo-peak analysis is depicted for $^{137}$Cs (at 662keV) using second order polynomial fitting for the background and Gaussian fitting for the photo-peak area, as illustrated in Fig.6.

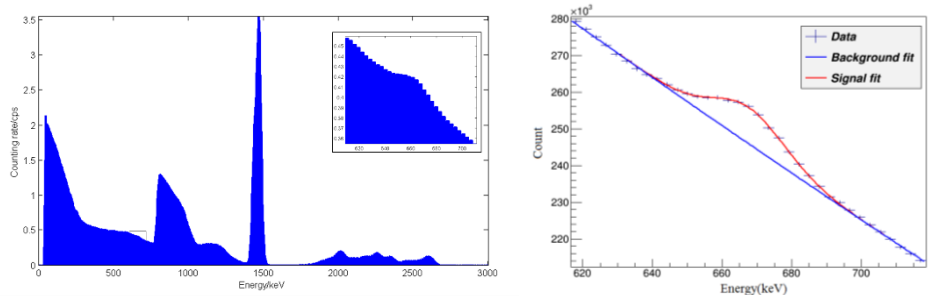

Fig. 6 In-situ spectrum for 7 days acquisition time

The detection efficiency calculated by experiment is (0.288±0.010) cps/(Bq/L), which is in good accordance with the MC simulation.

**6.Minimum detectable activity concentration**

A question occurs when measuring low activity materials that how to determine whether some kind of radionuclides are included. To solve the question, a physical quantity named MDAC is defined to describe the minimum detectable activity concentration of radionuclides for certain time. The MDAC of $^{137}$Cs in seawater tank was worked out as the following:

$$\text{MDAC} = \frac{4.65\sqrt{N_b}}{t I_\gamma \varepsilon_{sp} V}$$

According to the result of detection efficiency calculated by experiment, $I_\gamma \, \varepsilon_{sp} \cdot V$ is $(0.288 \pm 0.010)$ cps/(Bq/L). Then, the MDAC results of the prototype to 662keV photon from $^{137}$Cs are: 0.78Bq/L for 1 hour, 0.16Bq/L for 24 hours.

## 7. Conclusion

Conventional NaI(Tl) in-situ gamma-ray spectrometer for marine environmental monitoring generally had its MDAC about 1.0Bq/L. And in this paper, a prototype of LaBr$_3$:Ce in-situ gamma-ray spectrometer for marine environmental monitoring has been developed and applied for marine measurement. Because of its higher energy resolution than traditional NaI(Tl), a better MDAC of 0.78Bq/L is achieved. And the detect efficiency is $(0.288 \pm 0.01)$cps/(Bq/L), in good accordance with the Monte Carlo simulation.

The results showed that LaBr3:Ce would be a promising choice for the in-situ marine environmental monitoring, with better MDAC and nuclide identification. Meanwhile, the prototype is still in a preliminary stage, there are several aspects to be improved. The long-term stability is to be tested, and intrinsic background of the LaBr3:Ce material is also to be studied and discriminated in the future.

## Acknowledgements

This work is supported by the Ministry of Science and Technology special foundation work (2012FY130200), Tsinghua University, independent research project (2011080965), the National 863 Program (2012AA050907).